# Quantum and Post-Newtonian Effects in the Anomalistic Period and the Mean Motion of Celestial Bodies

Ioannis Haranas[1] Omiros Ragos[2] Ioannis Gkigkitzis[3] and Ilias Kotsireas[4]

[1,4] *Dept. of Physics and Computer Science, Wilfrid Laurier Science Building,*
*Room N2078, 75 University Ave. W. Waterloo, ON, N2L 3C5, Canada*
*e-mail: iharanas@wlu.ca, ikotsireas@wlu.ca*

[2] *Dept. of Mathematics, Faculty of Sciences, University of Patras, GR-26500 Patras, Greece*
*e-mail: ragos@math.upatras.gr*

[3]*Departments of Mathematics and Biomedical Physics, East Carolina University*
*124 Austin Building, East Fifth Street Greenville NC 27858-4353, USA*
*e-mail*: *gkigkitzisi@ecu.edu*

## Abstract

We study the motion of a secondary celestial body under the influence of the corrected gravitational force of a primary. We study the effect of quantum and relativistic corrections to the gravitational potential of a primary body acting on the orbiting body. More specifically, two equations are derived to approximate the perigee/perihelion/periastron time rate of change and its total variation over one revolution (i.e., the difference between the anomalistic period and the Keplerian period) under the influence of the quantum as well as post-Newtonian accelerations. Numerical results have been obtained for the artificial Earth satellite Molnya, Mercury, and, finally, the for the HW Vir c, planetary companion.

## 1 Introduction

The Newtonian potential energy that is usually considered to rule the motion of two bodies of mass $M_p$ (primary) and $m$ (secondary) which are separated by a distance $r$ is

$$V_N(r) = -\frac{GM_p m}{r} \qquad (1)$$

where $G$ is the Newtonian constant of gravitation. This potential is of course only approximately valid (e.g., Donoghue 1994). For large masses and/or large velocities, the General Relativity theory predicts that there exist relativistic corrections, which can be calculated and also verified experimentally (e.g., Bjorken and Drell 1964).

In the microscopic distance domain, we could expect that Quantum Mechanics would predict a correction to the gravitational potential in the same way that the radiative corrections of Quantum Electrodynamics lead to a modification of the Coulomb interaction (t'Hooft and Veltman 1974). Even though General Relativity constitutes a very well defined classical theory, it is not possible yet to combine it with Quantum Mechanics in order to create a satisfactory theory of Quantum Gravity. One of the basic obstacles that prevent this from happening is that general relativity does not actually fit the present paradigm for a fundamental theory, of a renormalizable quantum field theory. Gravitational fields can be successfully quantized on smooth-enough spacetimes (Capper et al. 1973), but the form of gravitational interactions is such that they induce unwanted divergences which cannot be absorbed by the renormalization of the parameters of the minimal General Relativity (Goroff and Sagnotti 1984). One can introduce new coupling constants and absorb the divergences, but, unfortunately, this leads to an infinite number of free parameters. Despite the difficulty described above, Quantum Gravity calculations can predict long distance quantum corrections. The main idea leading to quantum corrections at large distances is due to the interactions of massless particles which only involve their coupling energies at low energies, something that it is known from the GRT, even though at short distances the theory of quantum gravity differs, resulting to finite correction of order $O(G\hbar/c^3 r^3)$ where $\hbar$ is Planck's constant, and $c$ is the speed of light. The existence of a universal long distance quantum correction to the Newtonian potential should be relevant for a wide class of gravity theories. It is a well-known fact that the ultraviolet behaviour of Einstein's pure gravity can be improved, if higher derivative contributions to the action are added; in four dimensions they take the form (in usual notation)

$$\alpha R^{\kappa\lambda} R_{\kappa\lambda} + \beta R^2 \tag{2}$$

where $\alpha$ and $\beta$ are dimensionless coupling constants. What makes the difference is that the resulting classical and quantum corrections to gravity are expected to significantly alter the gravitational potential at short distances comparable to that of Planck length $\ell_p \approx 1.6 \times 10^{-35}$ m, but it should not really affect its behaviour at long distances. At long distances it is the structure of the Einstein-Hilbert action that actually determines that. At this point we should mention that some of the calculations of the corrections to the Newtonian gravitational potential result in the absence of a cosmological constant, which usually complicates the perturbative treatment to a significant degree because of the need to expand about a non flat background. In one-loop amplitude computation, one needs to calculate all first order corrections in $G$, which will include both the relativistic $O(G^2 M^2/c^2)$ and the quantum mechanical $O(G\hbar/c^3)$ corrections to the classical Newtonian potential (Hamber and Liu 1995).

The key ingredient that is leading to quantum corrections at large distances is the interactions of massless particles that only involve coupling at low energy. Both of these features are known from general relativity even if the full theory of quantum gravity is quite different at short distances. The action of gravity is determined by an invariance under general coordinate transformation and it is of the following form:

$$S = \int_V d^4x \sqrt{-g} \left( \frac{2R}{k^2} + \alpha R^2 + \beta R_{\mu\nu} R^{\mu\nu} + \gamma R_{\mu\nu} R^{\nu\eta} R^{\mu}_{\eta} \right) \tag{3}$$

We ignore the possibility of a cosmological constant, which experimentally must be very small. Here $R$ is the curvature scalar, $R_{\mu\nu}$ is the Ricci tensor, $g = \det g_{\mu\nu}$ and $g_{\mu\nu}$ is the metric tensor. Experiment determines $k^2 = 32\pi G$, where $G$ is Newton's constant, and $|\alpha|$, $|\beta| \leq 10^{74}$ (Stelle, 1978). The minimal general relativity consists of keeping only the first term, but higher powers of $R$ are not excluded by any known principle. The reason that the bounds on are so poor is that these terms have very little effect at low energies/long distance. The quantities $R$ and $R_{\mu\nu}$ involve two derivatives acting on the gravitational field (i.e., the metric $g_{\mu\nu}$). In an interaction each derivative becomes a factor of the momentum transfer involved, $q$, or of the inverse distance scale $q \approx \hbar/r$. We will say that $R$ is of order $q^2$. In contrast, $R^2$ or $R_{\mu\nu}R^{\mu\nu}$ are of order $q^4$. Thus, at small enough energies, terms of order $R^2$, $R^3$ etc. are negligible and we automatically reduce to only the minimal theory.

As a short digression on this theme, we note that Gutzwiller (1971, 1973, 1977) defined and studied a type of anisotropic Kepler problem with an essential goal: to identify links between classical and quantum mechanics (see also Gutzwiller 1990). The same model was resumed by Devaney (1978) and Casasayas and Llibre (1984), who went deeper into the problem. The anisotropic Manev problem, tackled by Craig et al. (1999), provided results that seem to build a bridge between classical mechanics, relativity, and quantum mechanics (as regards behavior in the neighborhood of collision). For important results, about the links between classical and quantum physics, we direct the reader to the paper of DeWitt-Morette (1979). Analogous results were obtained by Mioc et al. (2003) for the anisotropic Schwarzschild problem. Similarly, in Haranas et al. (2011) the authors investigate Yukawa-type potential effects in the anomalistic period of celestial bodies.

The main goal of this contribution is to use the acceleration resulting from the quantum and post Newtonian correction to the potential into the Gauss' planetary equations, to establish good approximations for the difference between the anomalistic period in this field and the Keplerian period. To this end, we resorted to the eccentric anomaly instead of the true anomaly, and thus calculate the changes in the anomalistic time after a full revolution. Next, we will also derive the effect of the quantum and post Newtonian correction the orbiting body's mean motion. Finally, our results will be validated using and Earth orbiting satellite, the planet Mercury, and the HR Vir-c system.

## 2. Corrections to the gravitational potential

According to Hamber and Liu (1995) and Haranas and Mioc (2009), and taking into account that $M_p >> m$ the corrected potential energy valid to order $G^2$, is:

$$V(r) = -\frac{GM_p m}{r}\left[1 - \frac{GM_p}{2c^2 r} - \frac{122G\hbar}{15\pi c^3 r^2}\right]. \tag{4}$$

The first correction, of order $GM/rc^2$, does not contain any power of $\hbar$, and is of the same form as various post-Newtonian corrections which we have dropped in taking the nonrelativistic limit (Weinberg, 1972). In fact, for a small test particle of mass $m$, this piece is the same as the expansion of the time component of the Schwarzschild metric, in other words

$$g_{00} = \frac{1 - 2GM_p/rc^2}{1 + 2GM_p/rc^2} \approx 1 - \frac{2GM_p}{rc^2}\left(1 - \frac{2GM_p}{rc^2}\right), \tag{5}$$

which is the origin of the static gravitational potential. One may see that two different length scales are embedded in the correction of the static Newtonian potential energy: The Planck length $\ell_p = \sqrt{\frac{G\hbar}{c^3}}$ and the Schwarzschild radius of the primary body $r_s = \frac{2GM_p}{c^2}$. Since these lengths are divided by the distance $r$ in the correction terms it must be taken into account that they are much smaller than $r$.

## 3 Equation for the passage time rate of change

Consider the unperturbed relative orbit of the secondary, obviously a Keplerian ellipse. Let $a$ be the semimajor axis, $e$ its eccentricity, $n$ its mean motion, and $M$ the mean anomaly. Then, $M$ is connected to the periastron passage time $T_0$ through the relation

$$M = n(t - T) \tag{6}$$

where $t$ denotes the time variable. The time rate of change of $T_0$ can be obtained by:

$$\frac{dT_0}{dt} = 1 + \frac{(t - T_0)}{n}\frac{dn}{dt} - \frac{1}{n}\frac{dM}{dt}. \tag{7}$$

Using Kepler's law, i.e. $GM = n^2 a^3$ we have that the time rate of change of the mean motion $n$ becomes:

$$\frac{dn}{dt} = -\frac{3n}{2a}\frac{da}{dt}. \tag{8}$$

Under the influence of any perturbing acceleration, the time derivatives of $M$ and $a$ can be found by Gauss' planetary equations (Vallado and McClain, 2007)

$$\frac{da}{dt} = \frac{2}{n\sqrt{1-e^2}}\left[eS\sin f + \frac{a(1-e^2)}{r}T\right] \tag{9}$$

$$\frac{dM}{dt} = n - \frac{(1-e^2)}{nae}\left[-\left(\cos f - \frac{2er}{a(1-e^2)}\right)R + \sin f\left(1 + \frac{r}{a(1-e^2)}\right)T\right] \tag{10}$$

where $S$ and $T$ are the radial and transverse components of this acceleration, and $f$ is the true anomaly. Given the corrections to the Newtonian potential in (4) we have that the corresponding force acting in the radial direction to be:

$$F(r) = -\frac{\partial V}{\partial r} = -\frac{GM_p m}{r^2} + \frac{G^2 M_p^2 m}{c^2 r^3} + \frac{122 G^2 M_p m\hbar}{5\pi c^3 r^4}. \tag{11}$$

First, we consider the quantum correction to be the only perturbing acceleration to the gravitational potential. This correction has only a radial acceleration component, which implies $T = 0$, and therefore we have that:

$$S_{qu} = \frac{122 G^2 M_p \hbar}{5\pi c^3 r^4} = \frac{122 G M_p \ell_p^2}{5\pi r^4} \qquad \text{m/s}^2 \tag{12}$$

By substituting Eqs. (8)-(9) and (10) into Eq. (7) we obtain the following equation for the corresponding time rate of change of the periastron time

$$\frac{dT_{0_{qu}}}{dt} = \left[ \frac{2r}{n^2 a^2} - \frac{(1-e^2)\cos f}{n^2 a e} - \frac{3e(t-T_0)\sin f}{na\sqrt{1-e^2}} \right] S_{qu} \tag{13}$$

This equation can be transformed in terms of the eccentric anomaly $E$. To this end, we use the following relations (Murray and Dermott, 1999):

$$r = a(1 - e\cos E) \tag{14}$$

$$\frac{dE}{dt} = \frac{n}{(1 - e\cos E)} \tag{15}$$

$$(t - T_0) = \frac{(E - e\sin E)}{n} \tag{16}$$

$$\cos f = \frac{\cos E - e}{1 - e\cos E} = \frac{a(\cos E - e)}{r} \tag{17}$$

$$\sin f = \frac{\sqrt{1-e^2}\sin E}{(1 - e\cos E)} = \frac{a\sqrt{1-e^2}\sin E}{r}. \tag{18}$$

Furthermore, using that $GM_p = n^2 a^3$ the equation for the time rate of the anomalistic time becomes:

$$\frac{dT_{0_{qu}}}{dE} = \frac{122\ell_P^2}{5\pi n a^2 (1-e\cos E)^3} \left[ 2(1-e\cos E) - \frac{(1-e^2)(\cos E - e)}{e(1-e\cos E)} - \frac{3e\sin E(E - e\sin E)}{(1-e\cos E)} \right] \tag{19}$$

The variation of $\Delta T_{0_{qu}}$ over one revolution is obtained by integrating $dT_{0qu}/dE$ from 0 to $2\pi$ and imposing that $e < 1$ we obtain:

$$\Delta T_{0_{qu}} = \frac{244}{5n(1-e)^3}\left(\frac{\ell_P}{a}\right)^2 \qquad \text{s/rev.} \tag{20}$$

We find that $\Delta T_{0_{qu}}$ scales as the square of the ratio of the Planck's length over the semimajor axis of the orbiting body. For real celestial body orbits, this relationship indicates that the quantum effect will be extremely small and, most possibly, not measurable by today's technology. Also, high eccentricity orbits Eq (20) result to a higher effect. In the case that the idea of quantized redshift proves to be valid, one might have to introduce a new cosmic quantum of action $\hbar_g = 6.322 \times 10^{67}$ J s (Haranas and Harney, 2009) and, therefore, a new cosmic Planck length $\ell^2_{Pl(cosmic)} = G\hbar_g/c^3 = 1.315 \times 10^{16}$ m. If this new quantum of action operates in Celestial Mechanics, it might affect large-scale phenomena.

Next, using Eq. (4) we examine the post Newtonian effect where the post-Newtonian correction is given by:

$$V_{PN} = \frac{G^2 M_p^2}{c^2 r^3} , \qquad (21)$$

Taking into account (14)-(16) as before, Eq. (20) is transformed in terms of the eccentric anomaly as follows:

$$\frac{dT_{0_{PN}}}{dE} = \frac{na^2}{c^2(1-e\cos E)^2}\left[2(1-e\cos E) - \frac{(1-e^2)(\cos E - e)}{e(1-e\cos E)} - \frac{3e\sin E(E - e\sin E)}{(1-e\cos E)}\right] \qquad (22)$$

Imposing that $0 < e < 1$ the integration of this equation with respect to the eccentric anomaly results in the following relation:

$$\Delta T_{0_{PN}} = \frac{3\pi n}{(e-1)^2}\left(\frac{a}{c}\right)^2 \qquad (23)$$

Comparing equations (20) and (23) we express the change of the anomalistic time due to post Newtonian effects in terms of the anomalistic time due to the quantum effects to be:

$$\Delta T_{0_{PN}} = \frac{15\pi(1-e)}{244} a^4 \left(\frac{n}{c\ell_p}\right)^2 \Delta T_{0_{qu}} . \qquad (24)$$

## 4. Mean Motion Change due to Quantum and Post-Newtonian Effects

In order to examine the corresponding change in the mean motion let us use Eq. (8) in which substitute Eq. (9) we obtain the following equation:

$$\frac{dn}{dt} = -\frac{3e\sin f}{a\sqrt{1-e^2}} R_{rad} . \qquad (25)$$

Next, using the same transformations in terms of the eccentric anomaly also making use of Eq. (12) Eq. (25) can be written in terms of the following way differential equation to be:

$$\frac{dn}{dE} = -\frac{366 ne\ell_p^2 \sin E}{5\pi a^2(1-e\cos E)^4} , \qquad (26)$$

separating and solving the differential equation with initial condition $n(0) = n_0 = \sqrt{GM/a^3}$ we obtain that:

$$n(E) = n_0 e^{\frac{122\ell_p^2}{5\pi a^2}\left[\frac{1}{(1-e\cos E)^3} - \frac{1}{(1-e)^3}\right]}, \qquad (27)$$

given that $122\ell_p^2/5\pi a^2 \ll 1$ Eq. (27) can be approximated as follows:

$$n(E) \cong n_0 \left[1 + \frac{122\ell_p^2}{5\pi a^2}\left(\frac{1}{(1-e\cos E)^3} - \frac{1}{(1-e)^3}\right)\right], \qquad (28)$$

This solution represents the mean anomaly $n$ as a function of the eccentric anomaly $E$ which satisfies the above initial condition. Similarly, assuming $e < 1$ taking the integral over one revolution of Eq. (26) we obtain that:

$$\Delta n_0 = -\frac{366 e n \ell_p^2}{5\pi a^2} \int_0^{2\pi} \frac{\sin E}{(1-e\cos E)^4} = 0. \tag{29}$$

And therefore over one revolution we obtain that the change to the initial mean motion $n_0$ becomes equal to:

$$n_{0_{final}} = n_0 + \Delta n_0 = \sqrt{\frac{GM_p}{a^3}}. \tag{30}$$

Next, we will proceed with the calculation of the change of the mean anomaly $n$ as a function of the eccentric anomaly $E$, and in the same way as before we obtain the following differential equation:

$$\frac{dn}{dE} = -\frac{3eG^2M_p^2 \sin E}{c^2 a^4 n(1-e\cos E)^3}, \tag{31}$$

Separating and solving we obtain that:

$$n(E) = \pm n_0 \sqrt{1 - \frac{3G^2 M_p^2}{n_0^2 a^4 c^2}\left[\frac{1}{(1-e\cos E)^2} - \frac{1}{(1-e)^2}\right]}, \tag{32}$$

since $3R_{Sch}/2a \ll 1$ Eq. (32) which can be approximated in the following way:

$$n(E) = \pm n_0 \left[1 - \frac{3R_{Sch}}{4a}\left[\frac{1}{(1-e\cos E)^2} - \frac{1}{(1-e)^2}\right]\right] \tag{33}$$

where $R_{Sch} = 2GM/c^2$ is the Schwarzschild radius which subjected to the initial condition $n(0) = n_0 = \sqrt{GM_p/a^3}$, and where the negative root will be considered to represent represents motion in the opposite way. Next taking the integral over one revolution of the second term of Eq. (31) that corresponds to the post Newtonian term we obtain that:

$$\Delta n_0 = \frac{3eG^2 M_p^2}{\pi c^2 a^4} \int_0^{2\pi} \frac{\sin E}{(1-e\cos E)^3} = 0, \tag{34}$$

and therefore over one revolution we obtain that the change to the initial mean motion $n_0$ becomes equal to:

$$n_{0_{final}} = n_0 + \Delta n_0 = \sqrt{\frac{GM_p}{a^3}}. \tag{35}$$

In figures 3 and 4 below we plot the functions $Q_{qu}(E) = (1-e\cos E)^{-3} - (1-e)^{-3}$ and $Q_{pN}(E) = (1-e\cos E)^{-2} - (1-e)^{-2}$ appearing in Eqs. (28) and (33). We find that for the fixed values of the orbital eccentricity $e = 0.0125, 0.100$ and $0.158$, $Q(E)$ obtains negative values as the eccentric anomaly goes through a full cycle. Moreover, higher eccentricities result to more negative $Q(E)$ values at corresponding higher range of $E$ values with the highest value of both $Q(E)$ at $E = 180°$. Using Eq. (28) we have that $n(E) = n_0\left(1 - \frac{122\ell_p^2}{5\pi a^2}Q_{qu}(E)\right) < n_0$ i.e. the quantum effects will reduce the value of the mean motion. Similarly, using Eq. (33) and since $Q(E)$ is always negative we have that $n(E) = \pm n_0 \left(1 + \frac{n_0 R_{Sch}}{4ea} Q_{pN}(E)\right) > n_0$, i.e. post Newtonian effects will increase the value of the mean motion. Next, we calculate the total period of such an orbit to be:

$$P_{tot} = \frac{2\pi}{n_{tot}} = \frac{2\pi}{n_0 + n_{qu} + n_{pN_1}} = \frac{2\pi}{n_0 \left[3 + \frac{122\ell_p^2}{5\pi a^2}\left(\frac{1}{(1-e\cos E)^3} - \frac{1}{(1-e)^3}\right) - \frac{3R_{Sch}}{4a}\left(\frac{1}{(1-e\cos E)^2} - \frac{1}{(1-e)^2}\right)\right]}, \quad (37)$$

which for $E = \pm 0, \pm 2\pi$ we obtain that $P_{tot} = 2\pi/3n_0 = P_{New}/3$. Similarly, taking account the negative sign of Eq. (33) the total period becomes:

$$P_{tot} = \frac{2\pi}{n_{tot}} = \frac{2\pi}{n_0 + n_{qu} + n_{pN_2}} = \frac{2\pi}{n_0 \left[1 + \frac{122\ell_p^2}{5\pi a^2}\left(\frac{1}{(1-e\cos E)^3} - \frac{1}{(1-e)^3}\right) + \frac{3R_{Sch}}{4a}\left(\frac{1}{(1-e\cos E)^2} - \frac{1}{(1-e)^2}\right)\right]}, \quad (38)$$

which for $E = \pm 0, \pm 2\pi$ we obtain that $P_{tot1} = 2\pi/n_0 = P_{New}$ where $P_{New}$ is the Newtonian period or the period due to the Newtonian potential. In a similar way we conclude that the Eq. (37) results to a reduction of the mean motion, where Eq. (38) in principle results to a larger or equal mean motion when compared to the Newtonian one. For example the mean motion of Mercury is $n_0 = 8.07 \times 10^{-7}$ rad/s and therefore $P = 2\pi/n_0 = 7.785 \times 10^6$ s, where using Eqs. (37) and (38) at $E = 0°, 45°, 90°, 270°, 360°$ we obtain that $P_{tot1} = 2\pi/n_0 = 2.595 \times 10^6$ s and also $P_{tot2} = 2\pi/n_0 = 7.785 \times 10^6$ s respectively.

## 5 Discussion and Numerical Results

We first calculate the quantum and relativistic effect on the perigee passage time of the earth orbiting artificial MOLNYA satellite with orbital parameters $a = 26554.30$ km, and $e = 0.7222$ and $n = 0.000145896$ rad/s (Capderou, 2005). These effects are:

$$\Delta T_{0_{qu}} = 5.779 \times 10^{-78} \quad \text{[s/rev]}, \tag{39}$$

$$\Delta T_{0_{PN}} = 1.396 \times 10^{-4} \quad \text{[s/rev]}. \tag{40}$$

Similarly, for Mercury, $a = 57909083$ km, $e = 0.205$ and $n = 8.07 \times 10^{-7}$ rad/s. Therefore the corresponding predicted variations of its perihelion passage time are:

$$\Delta T_{0_{qu}} = 9.373 \times 10^{-84} \quad \text{[s/rev]}, \tag{41}$$

$$\Delta T_{0_{PN}} = 0.448396 \quad \text{[s/rev]}. \tag{42}$$

Finally, for the HR Vir c planetary companion it is reported that (*http://exoplanet.eu/star.php?st=HW+Vir*) $a = 5.30 \text{ AU} = 7.95 \times 10^{11}$ m, $e = 0.45$, and therefore $n = 1.2 \times 10^{-8}$ rad/s, $M_{tot} = M_1 + M_2 = 0.485$ $M_S + 0.142$ $M_S = 1.247 \times 10^{30}$ kg, and thus we find that:

$$\Delta T_{0_{qu}} = 1.010 \times 10^{-83} \quad [\text{s/rev}], \tag{43}$$

$$\Delta T_{0_{PN}} = 2.62554 \quad [\text{s/rev}]. \tag{44}$$

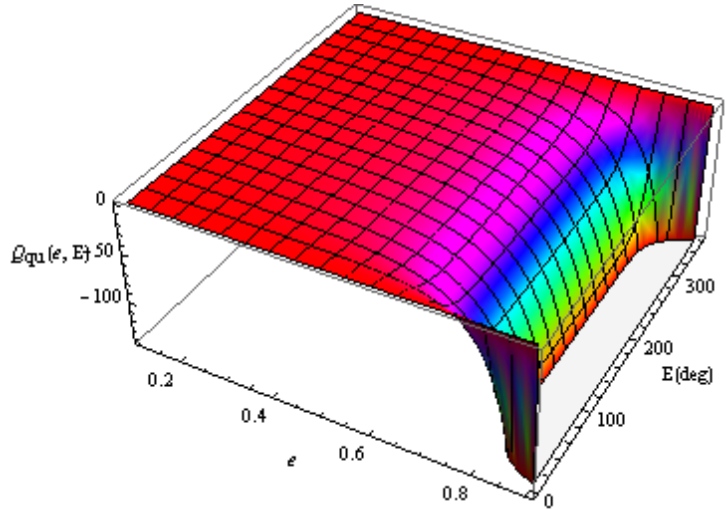

**Fig. 1** Three dimensional plot of quatnity $Q_{qu}(E)$ an a function of eccentric anomaly $E$ and orbital eccentricity $e$, in the range $0.11 \leq e \leq 0.9$

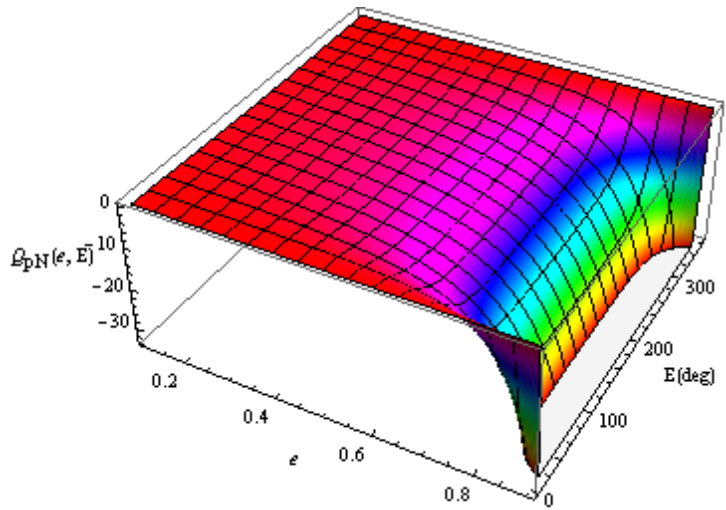

**Fig. 2** Three dimensional plot of quatnity $Q_{pN}(E)$ an a function of eccentric anomaly $E$ and orbital eccentricity $e$ in the range $0.11 \leq e \leq 0.9$.

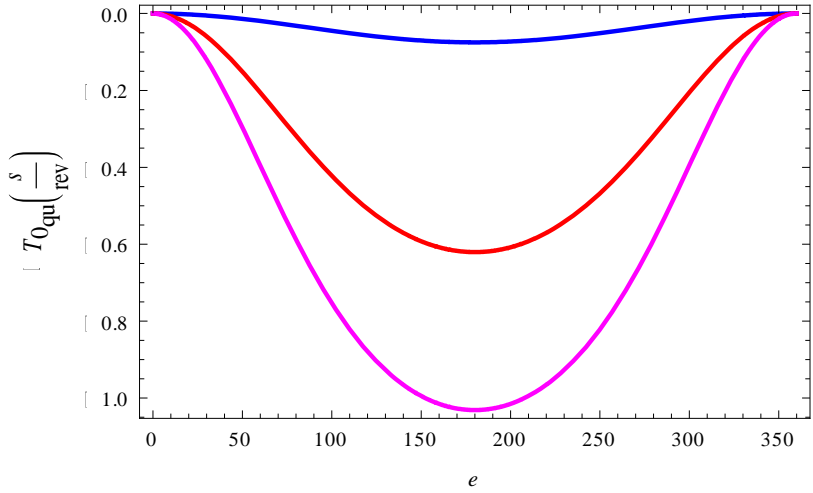

**Fig. 3** Plot of quatnity $Q_{qu}(E)$ an a function of eccentric anomaly $E$ for the following eccentricity values $e = 0.0125$ (red), $0.100$ (blue), $0.158$ (purple).

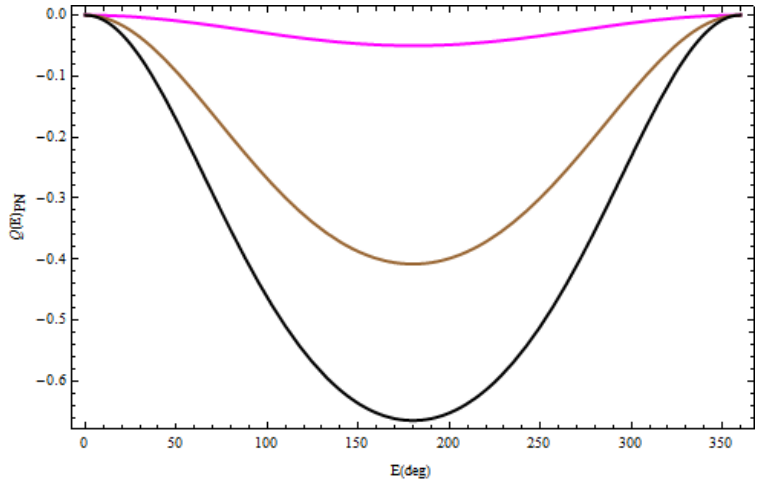

**Fig. 4** Plot of quatnity $Q_{PN}(E)$ an a function of eccentric anomaly $E$ for the following eccentricity values $e = 0.0125$ (red), $0.100$ (blue), $0.158$ (magenta).

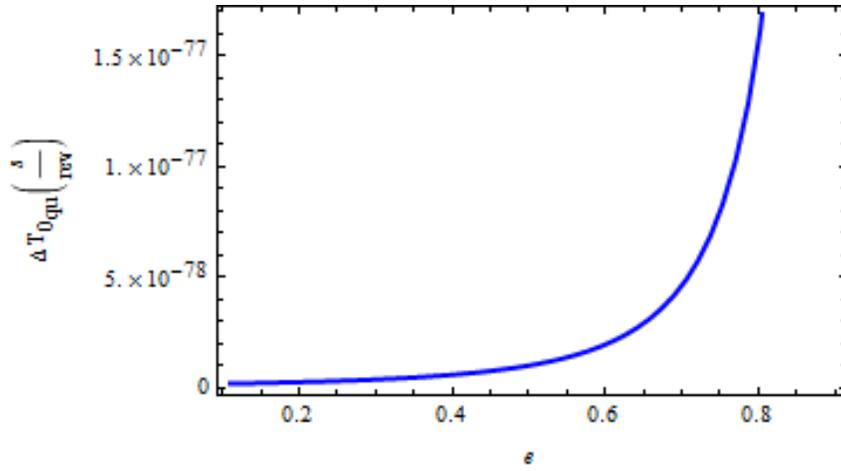

**Fig.5** Anomalistic time change per revolution due to quantum effect for Molnya satellite as a function of orbital eccentricity.

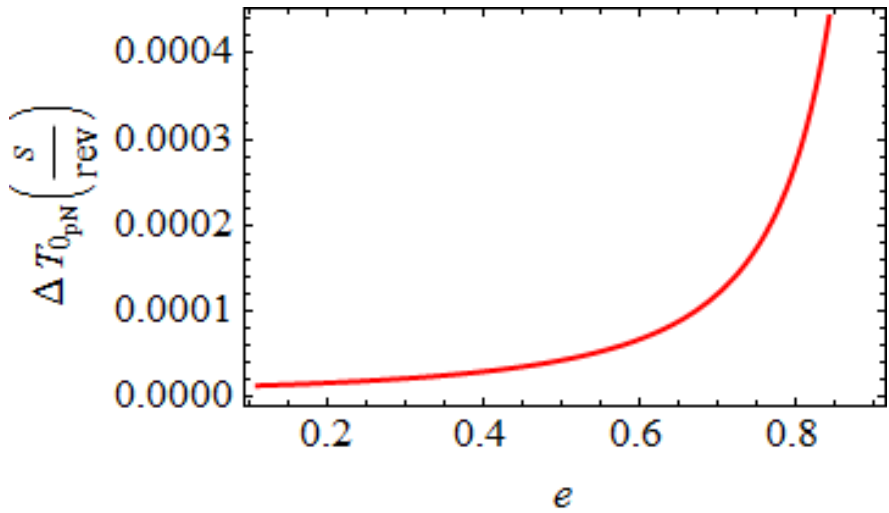

**Fig.6** Anomalistic time change per revolution due to post Newtonian effects for Molnya satellite as a function of the orbital eccentricity $e$.

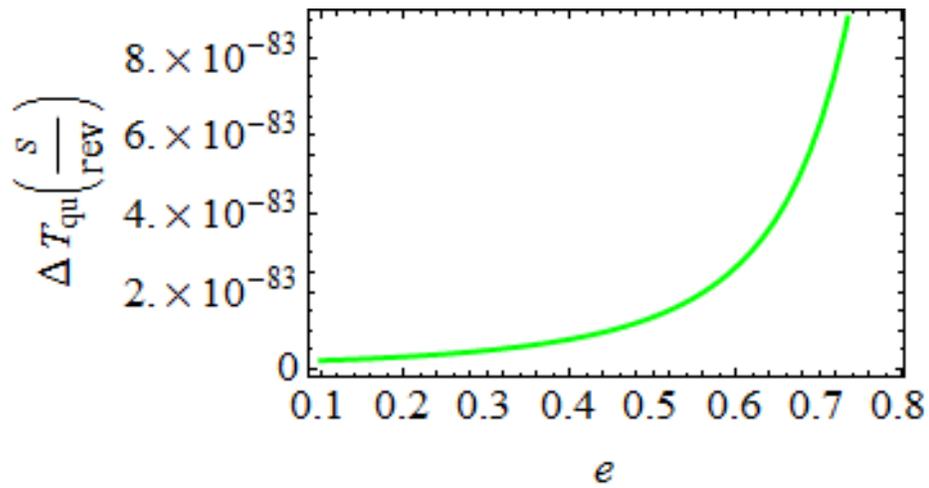

**Fig.7** Anomalistic time change due to quantum effect per revolution due for the HR Vir c planetary companion body as a function of orbital eccentricity $e$.

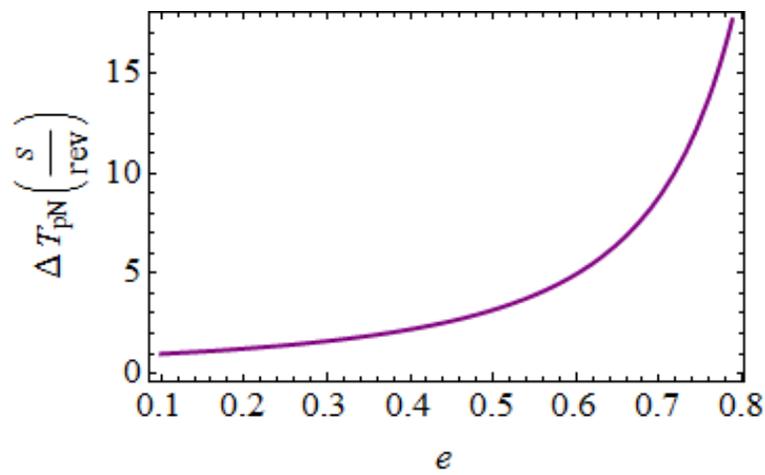

**Fig.8** Anomalistic time change due to post Newtonian effect per revolution for the HR Vir c planetary companion body as a function of orbital eccentricity $e$

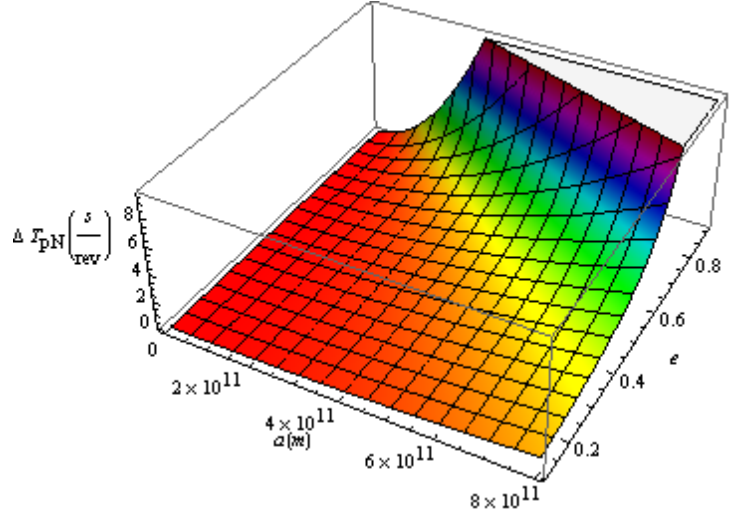

**Fig.9** Anomalistic time change due to post Newtonian effect per revolution for the HR Vir c planetary companion body as a function of orbital eccentricity *e* and semimajor axis *a*.

The plus signs of these variations indicate an advance of the perigee/perihelion/periastron passage times. Furthermore, in figure 2 we give a three dimensional plot of quatnity $Q_{qu}(E)$ as a function of eccentric anomaly $E$ and orbital eccentricity *e*. We find that as the eccentricity increases the value of $Q_{qu}(E)$ results to negative values in the full cycle of eccetric anomaly i.e. $0° \leq E \leq 360°$. In particular the function $Q_{qu}(E)$ drastically drops for very high eccentricity values in the range $0.73 \leq e \leq 0.9$. Figure 2 gives a three dimensional plot of quatnity $Q_{pN}(E)$ an a function of eccentric anomaly $E$ and orbital eccentricity *e*. In a similar way the function $Q_{pN}(E)$ drastically drops for very high eccentricity values in the range $0.79 \leq e \leq 0.9$. We find that for a full eccentric anomaly cycle the resulting values of $Q(E)$ satisfy $Q_{qu}(E) > Q_{pN}(E)$. Next, in figures 3 and 4 we plot the quantities $Q_{qu}(E)$ and $Q_{pN}(E)$ an a function of eccentric anomaly $E$ and for the following eccentricity values $e$ = 0.0125 (red), 0.100 (blue), 0.158 (purple). We find that higher eccentricity value results in more negative $Q$ values again satisfying $Q_{qu}(E) > Q_{pN}(E)$ and with a lowest $Q$ value occuring at $E = 180°$. In figures 5 and 6 we plot the anomalistic time change due to the quantum and post Newtonian effects as a function of orbital eccentricity the Molnya satellite. We find that as the orbital eccentricity of the secondary increases towards higher values, and in the case where $e \to 1$ the anomalistic time increases asymptotically. In figures 7 and 8 the anomalistic time change due to quantum effects and post Newtonian effects for the HR Vir planetary companion as a function of orbital eccentricity e. In figure 9, we compare quantum effects on the anomalistic time of planet Mercu to that of the HR Vir planetary companion. The calculated quantum corrections for Molnya, Mercury, and Vir-c are in the order of magnitude range $O(10^{-84})$s/rev $\leq \Delta T_{qu} \leq O(10^{-78})$s/rev, orders of magnitude that are not extremely small to be detected by today's technology. Similarly the post Newtonian effect range is $O(10^{-4})$ s/rev $\leq \Delta T_{qu} \leq O(10^{-1})$ s/rev. Even thought the two bodies (Mercury and Vir-c) are orbiting at a slightly different orbital distance, the mass of the primary is approximately the same i.e. $M_p \propto O(10^{30})$ might justify that the result is an order of magnitude apart for quantum correction effect. In relation to the quantum anomalistic time effects we find that the effect is larger for smaller semimajor axis orbits since it scales as $\Delta T_{0_{qu}} \propto 1/a^2$ thus being closer to the massive primary. Finally, the post Newtonian effect scales as $\Delta T_{0_{pN}} \propto a^2$ therefore a small semimajor axis results to a smaller anomalistic time change and vise versa.

## 6. Conclusions

The corrections to the gravitational potential due to Quantum Mechanics and the theory of General Relativity are consider for deriving equations that estimate the time rate of change and the variation over one revolution of the perigee/perihelion/periastron passage time $T_0$, of a secondary body orbiting a primary. We find that the quantum effects per revolution $\Delta T_{qu}$ are extremely small numbers to be detected by any of today's technology. Similarly, post Newtonian effects on the anomalistic time per revolution $\Delta T_{pN}$ lie in today's technology detectability range. Differences in anomalistic time can constitute an important way of testing and validating today's gravitational theories, by examining artificial satellite in highly elliptical orbits, planets as well as various exoplanetary stellar systems. Finally, post Newtonian parameter values can be obtain if precise measurements of anomalistic post Newtonian time effects are precisely measured.

## Acknowledgements

We would like to thank the associate editor of AS&SS Dr. Krzysztof Goździewski as well as an anonymous reviewer whose useful comments help improve this manuscript considerably.